\documentclass[prl,twocolumn,showpacs,showkeys,amsfonts,amssymb]{revtex4}
\usepackage{amsmath}
\usepackage{graphics}
\usepackage[pdftex]{graphicx}
\usepackage{graphicx}

\begin{document}
\title{Translocation of a Polymer through a Nanopore across a Viscosity Gradient}
\author{Hendrick W. de Haan, Gary W. Slater}
\date{\today}
\affiliation{Physics Department, University of Ottawa, Ottawa, Ontario, Canada, K1N 6N5}
\begin{abstract}

The translocation of a polymer through a pore in a membrane separating fluids of different viscosities is studied via several computational approaches.
Starting with the polymer halfway, we find that as a viscosity difference across the pore is introduced, translocation will predominately occur towards one side of the membrane.
These results suggest an intrinsic pumping mechanism for translocation across cell walls which could arise whenever the fluid across the membrane is inhomogeneous.
Somewhat surprisingly, the sign of the preferred direction of translocation is found to be strongly dependent on the simulation algorithm:
for Langevin Dynamics (LD) simulations, a bias towards the \emph{low} viscosity side is found while for Brownian Dynamics (BD), a bias towards the \emph{high} viscosity is found.
Examining the translocation dynamics in detail across a wide range of viscosity gradients and developing a simple force model to estimate the magnitude of the bias, the LD results are demonstrated to be more physically realistic.
The LD results are also compared to those generated from a simple, one dimensional random walk model of translocation to investigate the role of the internal degrees of freedom of the polymer and the entropic barrier.
To conclude, the scaling of the results across different polymer lengths demonstrates the saturation of the preferential direction with polymer length and the non-trivial location of the maximum in the exponent corresponding to the scaling of the translocation time with polymer length.

\end{abstract}
\maketitle

\section{Introduction}

The transport of polymeric molecules from one space to another through a constricting passage represents a fundamental process with a diverse set of applications.
Within biological systems, there are numerous examples that fall under this umbrella 
including the transport of DNA, RNA, and proteins across cell membranes \cite{albe89},
the packing of DNA or RNA into - and subsequent release from - viral capsids \cite{black89},
and the passage of proteins across the eukaryotic endoplasmic reticulum \cite{rapoport2007}. 
In addition to these naturally occurring examples, there is considerable interest in this topic due to the implications for the design of nanofluidic devices.
By far the most prominent of such applications concerns using the translocation of DNA through a synthetic nanopore as the basis for rapid and cheap sequencing technologies \cite{branton2008}.

As a consequence, there has been a large volume of research on this topic including many theoretical and simulation studies (\emph{c.f.} \cite{muthubook} and references therein).
In the majority of these studies, the conditions across the pore are taken to be equivalent with the exception that, for the case of driven translocation, there is a driving force towards one side.
That is, modelling parameters such as the quality of the solvent and the membrane-polymer interactions are the same on both sides.
However, for both biological and synthetic applications, these conditions are unlikely to be uniform across the membrane.
For example, considering translocation across a cell wall, the type and concentration of solutes is not equivalent for the intracellular and extracellular fluid
and, as an associated effect, the viscosity of the fluid is different across the membrane.
While the majority of the literature neglects such effects and considers translocation driven by an external field,
there have been a number of studies considering a bias which results from an asymmetry across the membrane.
Examples include a bias arising from varying solvent conditions across the pore \cite{wie2007,kapahnke2010}, 
differing concentrations of obstacles across the membrane \cite{gopinathan2007},
chaperone assisted translocation \cite{ambjornsson2004,yu2011}
and translocation induced by the adsorption of monomers to one side of the membrane \cite{park1998,milchev2004}.

In this work, we study translocation in the presence of a viscosity gradient where the viscosity on the $cis$ side is different than the viscosity on the $trans$ side.
Performing both Langevin Dynamics (LD) and Brownian Dynamics (BD), we find that a viscosity gradient introduces a bias to monomers at the interface.
Starting with the polymer halfway through the pore, this bias establishes a preferred direction for translocation which grows with an increasing discrepancy between viscosities.
In what may be a surprising result, which side is preferred depends on the simulation algorithm:
in LD, more events occur to the \emph{low} viscosity side; in BD more events occur to the \emph{high} viscosity side.
To characterize this drastic difference of results, simulations exploring the details of the dynamics are performed.
A simple force model is also employed to estimate the magnitude of the viscosity gradient bias for both LD and BD results.
From this analysis, we take the LD results to be more physical and additional simulations at different polymer lengths $N$ are performed 
to examine the dependence of the scaling of both the preferential direction and translocation time $\tau$ on the magnitude of the viscosity gradient.
We find that the strength of the preferential direction slowly grows with $N$.
Comparing these results to those obtained for a single random walker at a viscosity interface indicates that this $N$ dependence arises from the internal degrees of freedom of the polymer.
For the translocation time, the scaling exponent $\alpha$ obtained from $\tau \sim N^\alpha$ is found to decrease as the bias at the interface increases, as expected.
However, the maximum in $\alpha$ is shown to occur not when the viscosities across the pore are equal, but rather when the viscosity on one side is slightly higher than the other.
The competing mechanisms which lead to this result are discussed.

\section{Simulation Setup}

To model the polymer, we employ an approach that is by now common for coarse-grained simulations \cite{slater2009}.
To prevent overlap of the monomers, a shifted and truncated Lennard-Jones potential (often called the WCA potential \cite{weeks1978}) is used for excluded volume interactions.
Defining $\epsilon$ to be the energy of the interaction and $\sigma$ to be the monomer diameter, the potential as a function of the centre to centre distance between monomers, $r$, is given by
\begin{eqnarray}
 U_{WCA} (r) =
 \begin{cases} 
  4 \epsilon \left[ \left( \frac{\sigma}{r}\right)^{12} - \left( \frac{\sigma}{r} \right)^6 \right] + \epsilon         & \text{for } r  <  r_c  \\
  0                                                                & \text{for } r \geq r_c
 \end{cases}
\label{WCA}
\end{eqnarray}
where $r_c = 2^{1/6} \sigma$ is the cut-off distance.
Bonds between monomers along the polymer are modelled by the FENE potential which is given by 
\begin{equation}
U_{\textrm{FENE}}(r) = - \frac{1}{2} k r^2_0 \textrm{ln} \left( 1 - \frac{r^2}{r_0^2} \right).
\end{equation}
Following the model of Kremer and Grest \cite{grest1986} we set $k=30 \epsilon/\sigma^2$ and $r_0 = 1.5 \sigma$.
A continuous surface is used to model the nanopore-containing membrane.
To achieve a pore that is as large as possible while still ensuring a single file process, the radius of the pore is set to 1.5$\sigma$ \cite{dehaan2010}.

In this work, an explicit solvent is not included and instead the effects of the solvent are included implicitly in the equation of motion for the monomers.
While this approach neglects hydrodynamic interactions, the computational savings are significant and allow for a much more thorough investigation.
Two simulation approaches with implicit solvent are employed: Langevin Dynamics (LD) and Brownian Dynamics (BD).
In LD, the effects of a solvent are included implicitly by adding two terms to the equation of motion.
First, a drag force $F_f$ proportional to the monomer velocity $\vec{v}$ represents the dissipation aspects of monomer-solvent interactions.
Second, a random force models the fluctuations.
The final equation of motion is given by:
\begin{equation}
m\ddot{\vec{r}} = -\nabla U(\vec{r}) - \zeta \vec{v} + \vec{R}(t),
\label{LD}
\end{equation}
where $m$ is the mass of the monomer, $U(\vec{r})$ is sum of the conservative potentials, $\zeta$ is the friction coefficient, and $\vec{R}(t)$ is the random term.
To ensure that the fluctuation-dissipation theorem is obeyed, $\vec{R}(t)$ satisfies the following criteria:
\begin{eqnarray}
\langle \vec{R}_i(t) \rangle & = & 0 \\
\langle \vec{R}_i(0) \cdot \vec{R}_j(t) \rangle & = & 2 k_B T \zeta \delta(t) \delta_{ij}
\end{eqnarray}
where the subscript $i,j$ denote components along cartesian coordinates.

Considering the dissipation term, the drag on the particle moving through a fluid is given by Stoke's relation
\begin{equation}
\vec{F}_f = -6 \pi \eta R_p \vec{v},
\end{equation}
where $R_p$ is the radius of the particle.
The friction coefficient is thus proportional to the viscosity, $\zeta = 6 \pi \eta R_p$,
and considering a fixed particle size, varying $\zeta$ thus directly corresponds to varying the viscosity of the fluid.
To make this identification, $\zeta$ has units of $\sqrt{\epsilon m/\sigma^2}$
and we thus define a dimensionless viscosity given by
\begin{equation}
\tilde{\eta} = \frac{\zeta}{\sqrt{\epsilon m/\sigma^2}}.
\end{equation}

In BD, the dynamics are taken to be in the overdamped limit where the inertia of a particle is negligible compared to the drag and random forces.
A particle thus immediately reaches terminal velocity and the trajectory consists of a series of uncorrelated jumps induced by the random force.
In these simulations, Eqn. \ref{LD} is modified by dropping the inertial term to yield:
\begin{equation}
\zeta \vec{v} = -\nabla U(\vec{r}) - \vec{R}(t).
\label{eqn:BD}
\end{equation}
Performing both LD and BD, the same model for the polymer and nanopore is used between simulations.
However, in BD, the WCA potential is capped at 75 $\epsilon/\sigma$ to prevent breaking of the bonds due to large overlaps of monomers arising from large jumps.

\begin{figure}[h]
 	\centering
	\includegraphics[width=0.50\textwidth]{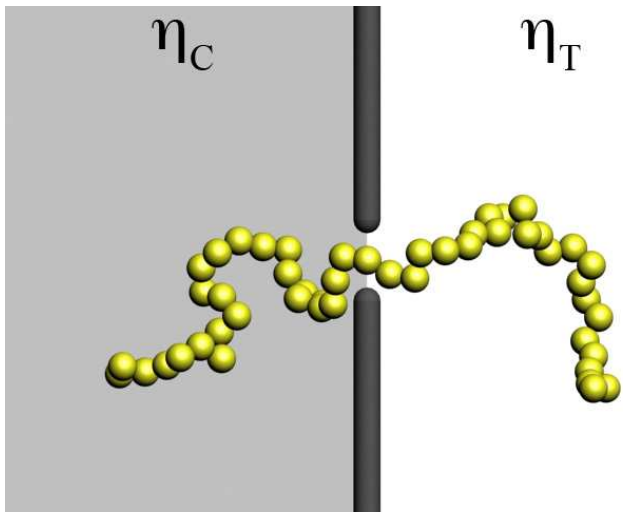} 
	\caption{Schematic of unbiased translocation across a viscosity gradient.  The viscosity on the $cis$ side is held fixed at $\eta_C$ = 1.0 while the viscosity on the $trans$ side is varied from $\eta_T$=0.1 to $\eta_T$=10.}
	\label{fig:scheme}
\end{figure}

The system is constructed by having different viscosities of fluid on either side of the membrane.
As shown in Fig. \ref{fig:scheme}, the interface is placed halfway through the pore.
Particles crossing the surface are treated as point particles such that the friction coefficient in eq. \ref{LD} is chosen based on which side of the interface the centre of the monomer is on.
As we study unbiased translocation, simulations begin with the polymer halfway through the pore with an equal number of monomers on \textit{cis} and \textit{trans}.
The middle monomer, which is considered to be on $cis$ at $t=0$, is fixed to allow the polymer to equilibrate.
Following equilibration, the polymer is released and the direction and time of translocation are recorded.
For all simulations, the viscosity on the \textit{cis} side is held at $\tilde{\eta}_C = 1.0$.
The viscosity on \textit{trans} is varied from $\tilde{\eta}_T = 0.1$ to 9.0.

For the LD simulations, polymers of length $N = 25, 49, 75, 99$ are studied.
In the BD simulations, a polymer of length $N=49$ was simulated.

\section{Results}

\subsection{Langevin Dynamics}

\begin{figure}[h]
 	\centering
	\includegraphics[width=0.50\textwidth]{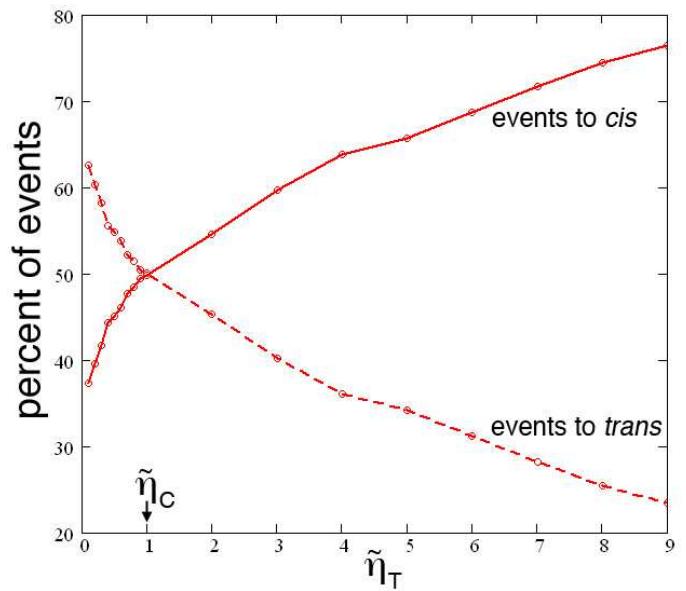} 
	\caption{Fraction of events to the $\textit{cis}$ side (solid) and $\textit{trans}$ side (dashed) as a function of $\tilde{\eta}_T$ for LD simulations with a polymer of size $N$=49.
	The low viscosity side is always preferred.}
	\label{fig:LD_PD}
\end{figure}

Beginning with the results from LD simulations, the data for the preferential direction is displayed in Fig. \ref{fig:LD_PD}.
For $\tilde{\eta}_T < \tilde{\eta}_C$, more events occur towards \textit{trans}.
Conversely, for  $\tilde{\eta}_T > \tilde{\eta}_C$, more events occur towards \textit{cis}.
Hence, the low viscosity side is always preferred (with equal probability at $\tilde{\eta}_T = \tilde{\eta}_C$ as required.
Further the strength of the preferential direction increases with an increasing viscosity gradient;
for  $\tilde{\eta}_T/\tilde{\eta}_C = 9$, nearly 80\% of all events end up on the $\textit{cis}$ wall.

\begin{figure}[h]
 	\centering
	\includegraphics[width=0.50\textwidth]{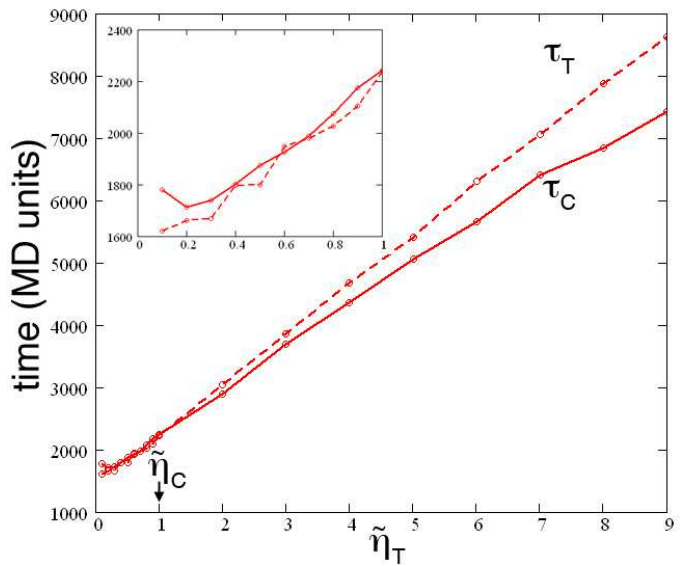} 
	\caption{Translocation time to the $\textit{cis}$ side (solid) and $\textit{trans}$ side (dashed) as a function of $\tilde{\eta}_T$ for LD simulations for a polymer of size $N$=49.
	The inset highlights the behaviour at low viscosities.}
	\label{fig:LD_tau}
\end{figure}

The respective translocation times are shown in Fig. \ref{fig:LD_tau}.
Unsurprisingly, the translocation time to the low viscosity is always less than the time to the high viscosity side.
However, the difference between $\tau_T$ and $\tau_C$ is small compared to the variation of both with $\eta_T$.
That is, although increasing $\tilde{\eta}_T$ introduces a preferred direction towards $\textit{cis}$, the net time increases due to the slower dynamics on the $\textit{trans}$ side.

\subsection{Brownian Dynamics}

\begin{figure}[h]
 	\centering
	\includegraphics[width=0.50\textwidth]{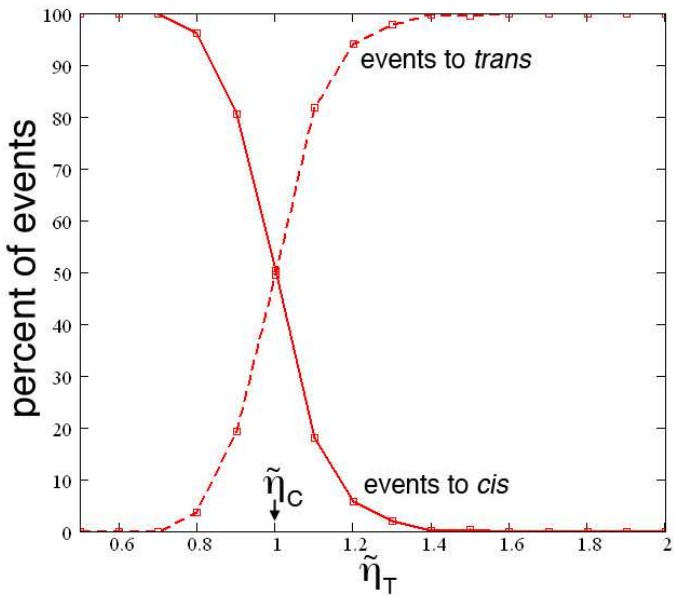} 
	\caption{Percent of events to the $\textit{cis}$ side (solid) and $\textit{trans}$ side (dashed) as a function of $\tilde{\eta}_T$ for BD simulations for polymers of size $N$=49.
	The high viscosity side is always preferred.}
	\label{fig:BD_PD}
\end{figure}

The results for the preferential direction obtained via BD simulations are shown in Fig. \ref{fig:BD_PD}.
In exact disagreement with the LD results, the high viscosity result is always preferred in BD.
Further, the strength of the preferential direction grows much faster with $\tilde{\eta}_T$ than in the LD results.
Here, when $\tilde{\eta}_T/\tilde{\eta}_C = 2$, essentially all events end up on the \textit{trans} wall.

\begin{figure}[h]
 	\centering
	\includegraphics[width=0.50\textwidth]{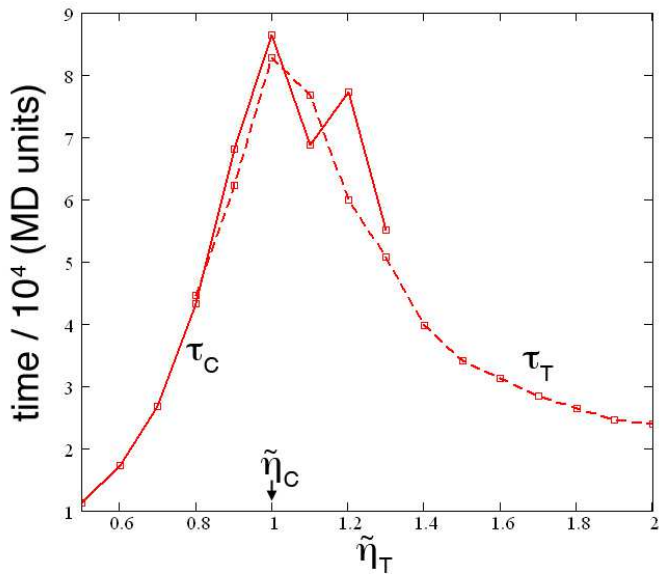} 
	\caption{Translocation time to the $\textit{cis}$ side (solid) and $\textit{trans}$ side (dashed) as a function of $\tilde{\eta}_T$ for BD simulations at a polymer size of $N$=49.
	Data is shown only when at least 10 events are recorded.}
	\label{fig:BD_tau}
\end{figure}

The corresponding translocation times are shown in Fig. \ref{fig:BD_tau}.
Again, significant disagreement with the LD results is found.
In LD, both $\tau_T$ and $\tau_C$ increased monotonically with increasing $\tilde{\eta}_T$.
In the BD results, $\tau_T$ and $\tau_C$ are maximum in the vicinity of $\tilde{\eta}_T/\tilde{\eta}_C = 1$ and the translocation times on either side decrease with an increasing viscosity difference.

\subsection{LD vs. BD}

\subsubsection{Diffusion in an inhomogeneous medium}

It is worth discussing the marked difference between the LD and BD results.
We first note that both results follow from work we have recently published studying a single random walker at a viscosity interface \cite{dehaan2012c}.
In that work, it was shown that for BD simulations, particles near the interface will tend to accumulate on the high viscosity side
while for LD simulations, the particles preferentially end up on the low viscosity side.

The difference fundamentally amounts to the extent to which the particle ``feels" the interface as it is crossing it.
In BD, there is no memory in the system; the dynamics at each time step are independent of the dynamics at all previous time steps.
Correspondingly, a particle that is in the low viscosity side will jump over the interface - with a jump length given by the low viscosity - and land in the high viscosity side.
Thus, a particle going from low to high viscosity will jump far into the high viscosity side.
Conversely, a particle jumping from high to low viscosity regions will do so with a short jump length and land near to the interface.
The net effect is that there is a bias at the interface favouring particles moving to the high viscosity side.
Given the independent dynamics, the particle does not see the interface as it is crossing it - it simply lands in a region of different viscosity.

In LD simulations however, there is memory in the system due to the inertial term in the equation of motion.
That is, there is a finite correlation time in the system and it takes a measurable amount of time for the velocity of the particle at any given time to be washed out.
Correspondingly, as a particle is crossing the interface, it can feel the interface since the rate at which the particle inertia is being damped will change.
This means that particles crossing from low to high will be stopped short (compared to dynamics in the low side) 
while particles crossing from high to low will land further past the interface (compared to dynamics in the high side).
This difference amounts to an effective bias at the interface favouring the low viscosity side.
Hence, in BD, particles tend to accumulate on the high viscosity side while in LD, there is a bias favouring the low viscosity side  \cite{dehaan2012c}.
To be more precise, BD corresponds to an Ito formulation and LD corresponds to an isothermal formulation for the problem of diffusion in an inhomogeneous medium \cite{lancon2002,sokolov2010}.

\begin{figure}[h]
 	\centering
	\includegraphics[width=0.45\textwidth]{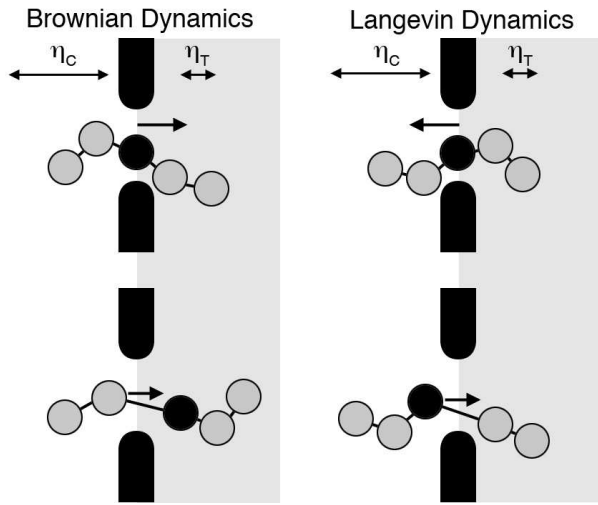} 
	\caption{Schematic for the dynamics of monomers crossing the viscosity interface for BD and LD.}
	\label{fig:schemebdld}
\end{figure}

Applying these ideas to polymer translocation gives context to the contradictory results shown above.
In LD, monomers at the interface will tend towards the low viscosity side and, correspondingly, a net bias to the low viscosity side is established yielding the preferential directions shown in \ref{fig:LD_PD}.
For the BD results, particles tend to accumulate on the high side and translocation to the high side is the most likely outcome.
A comparison between the BD and LD dynamics of a particle at the interface is shown in Fig. \ref{fig:schemebdld}.
In both cases, we consider a monomer at the interface jumping in the preferred direction (high viscosity for BD, low viscosity for LD).
The jump to the right or left yields an increase in chain tension with respect to the monomers left behind.
In BD, the particle moves slower once it is on the high viscosity side and it is effectively trapped.
To resolve the increased tension, the chain on the low viscosity side - which moves and relaxes more freely - will be pulled towards the high viscosity side.
Hence, for BD, both the bias and resulting chain tension favour translocation to the high viscosity side and this results in the rapid establishment of a preferential direction as seen in Fig. \ref{fig:BD_PD}.
In LD, the particle jumps to the low viscosity side.
However, as the monomers remaining in the high viscosity move slowly, there will be increased tension tending to bring the monomer back to the interface.
For LD, the bias and resulting chain tension thus work against each other and, as seen in Fig. \ref{fig:LD_PD}, the establishment of a preferential direction is much more gradual than in BD.

\subsubsection{Measuring the dynamics}

To get insight into the different behaviour for the translocation times, we use the physical picture outlined in Fig. \ref{fig:schemebdld} to develop a simple test.
Roughly speaking, there are two factors which affect the translocation time: the amount of diffusive motion and the time scale of the dynamics.
The establishment of a bias at the interface would be expected to reduce the amount of diffusive motion and thus speed up translocation.
Conversely, if the bias is generated by increasing the viscosity on one side of the membrane, the reduced rate of dynamics would be expected to increase the translocation time.
To investigate these opposing contributions for both BD and LD, simulations were performed were the translocation coordinate $s$ was recorded at each time step.
Here, $s$ is defined to be the number of monomers on $trans$ or, equivalently, the monomer currently in the pore where the numeration begins with $s=0$ at the end of the polymer on the $trans$ side.
Taking a specific monomer, denoted $s^*$, two tests were performed.
First, the number of times that the polymer is displaced by a distance $\Delta s$ was recorded by measuring the number of times that monomers further along the chain $s^* + \Delta s$ and back along the chain $s^* - \Delta s$ were in the pore.
Second, the time required to achieve the displacement was recorded.
The first test thus indicates the amount of diffusive motion as it indicates the number of times the polymer travels back and forth through the pore.
The second test measures the time scale of the dynamics.

\begin{figure}[h]
 	\centering
	\includegraphics[width=0.45\textwidth]{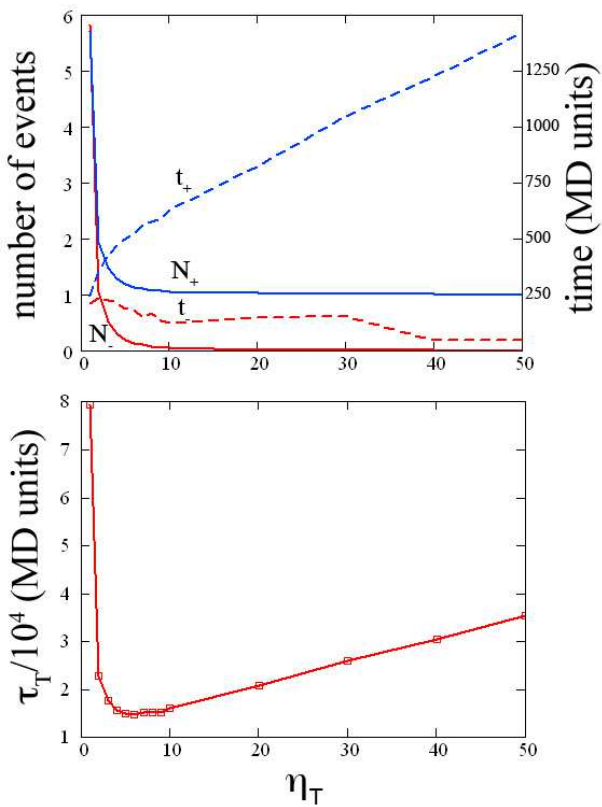} 
	\caption{Dynamics of monomers at the interface for BD for a polymer of size $N=49$.
	a) displays the number of displacements forward $N_+$ and backwards $N_-$ along with the average times for these events given by $t_+$ and $t_-$.
	b) displays the average net $trans$ translocation time.}
	\label{fig:BD_dyns}
\end{figure}

Note that as noise is directly added to the equation of motion for each monomer, there will always be significant local fluctuations.
As we are interested in the dynamics on a large time/length scale, a slightly larger displacement is chosen and we set $\Delta s = 3$.
The polymer length is set to 49 and we concentrate on the results for $\tilde{\eta}_T > \tilde{\eta}_C$.
Further, only trajectories ending at the preferred side are considered (in fact, as much of this work will be done at a 100\% preferential direction, this is a necessity since events occur $only$ in one direction).
For BD, this means events occurring to $trans$ while for LD this means events to $cis$.
In either case, the displacements in the direction of the preferential direction are considered positive while those in the other direction are negative.
For similar reasons, calculations are performed only for the monomers initially on the opposite side of the membrane as it is necessary for these monomers to pass through the pore in order for translocation to be achieved.
Hence, for BD, the number of events forward and backward are averaged over $s^*=25$ to $s^*=46$; for LD over $s^*=23$ to $s^*=2$.
For each $s^*$, the number of times that $s=s^* + 3$ is recorded as a $N_+$ for BD (and $N_-$ for LD); the number of times that $s=s^* - 3$ is recorded as a $N_-$ for BD (and $N_+$ for LD).
Associated times for these events are given by $t_+$ and $t_-$.
Results for the BD simulations are shown in Fig. \ref{fig:BD_dyns}.

In Fig. \ref{fig:BD_dyns} a, note the rapid decrease in $N_+$ and $N_-$ as $\eta_T$ increases.
By $\eta_T=10$, the values have saturated at $N_+ =1$ and $N_-=0$.
That is, the polymer never goes backwards (at least as far as 3 monomers) and goes forwards only once: there is no diffusive motion at the interface.
While the timescale for the forward motion $t_+$ is increasing with $\eta_T$ as expected, the reduction of $N_+$ is dominant and, as shown in Fig. \ref{fig:BD_dyns} b,
the translocation time plummets to less than 1/4 of its initial value by $\eta_T=10$.
In fact, by $\eta_T=10$, $\tau_T$ has started to increase as the reductions in $N_+, N_-$ have saturated while $t_+$ continues to increase.

\begin{figure}[h]
 	\centering
	\includegraphics[width=0.45\textwidth]{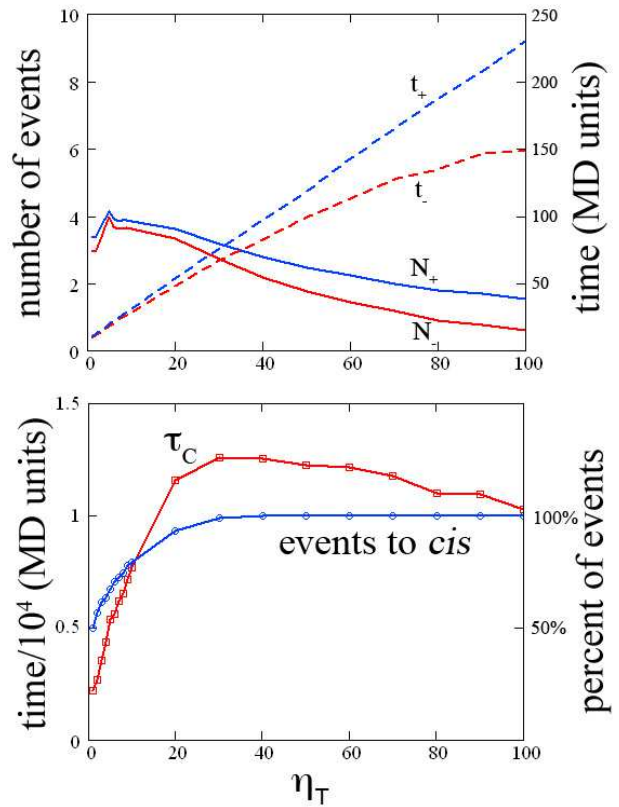} 
	\caption{Dynamics of monomers at the interface for LD for a polymer of size $N=49$.
	a) displays the number of displacements forward $N_+$ and backwards $N_-$ along with the average times for these events given by $t_+$ and $t_-$.
	b) displays the average net translocation time and the number of events occurring towards the low viscosity side.}
	\label{fig:LD_dyns}
\end{figure}

The corresponding results for LD are shown in Fig. \ref{fig:LD_dyns}.
In Fig. \ref{fig:LD_dyns} a, contrary to the BD results, $N_+$ and $N_-$ initially increase.
A physical picture for this result is given in Fig. \ref{fig:schemebdld}.
Although monomers tend towards the low viscosity side, they are pulled back from the slow moving chain on the high viscosity side yielding increased fluctuations.
There is agreement with the BD results in that $t_+$ increases with increasing $\eta_T$ as expected.
Consequently, as both $N_+$ and $t_+$ increase, the net translocation time initially increases as shown in Fig. \ref{fig:LD_dyns} b.
The initial increase in $N_+$ is short lived and, after $\eta_T \approx 5$, $N_+$ decays with increasing $\eta_T$ in agreement with the BD results.
However, contrary to the BD results, the decay of $N_+$ is weaker than the increase in $t_+$ and the translocation time continues to increase.
At even higher viscosities, the balance of these factors reverses and after $\eta_T = 30$, $\tau_C$ slowly decreases with increasing $\eta_T$.
As shown in Fig. \ref{fig:LD_dyns} b, the peak in $\tau_C$ coincides with the preferential direction plateauing at 100\%.

\subsubsection{Estimating the bias due to the viscosity gradient}

\begin{figure}[h]
 	\centering
	\includegraphics[width=0.50\textwidth]{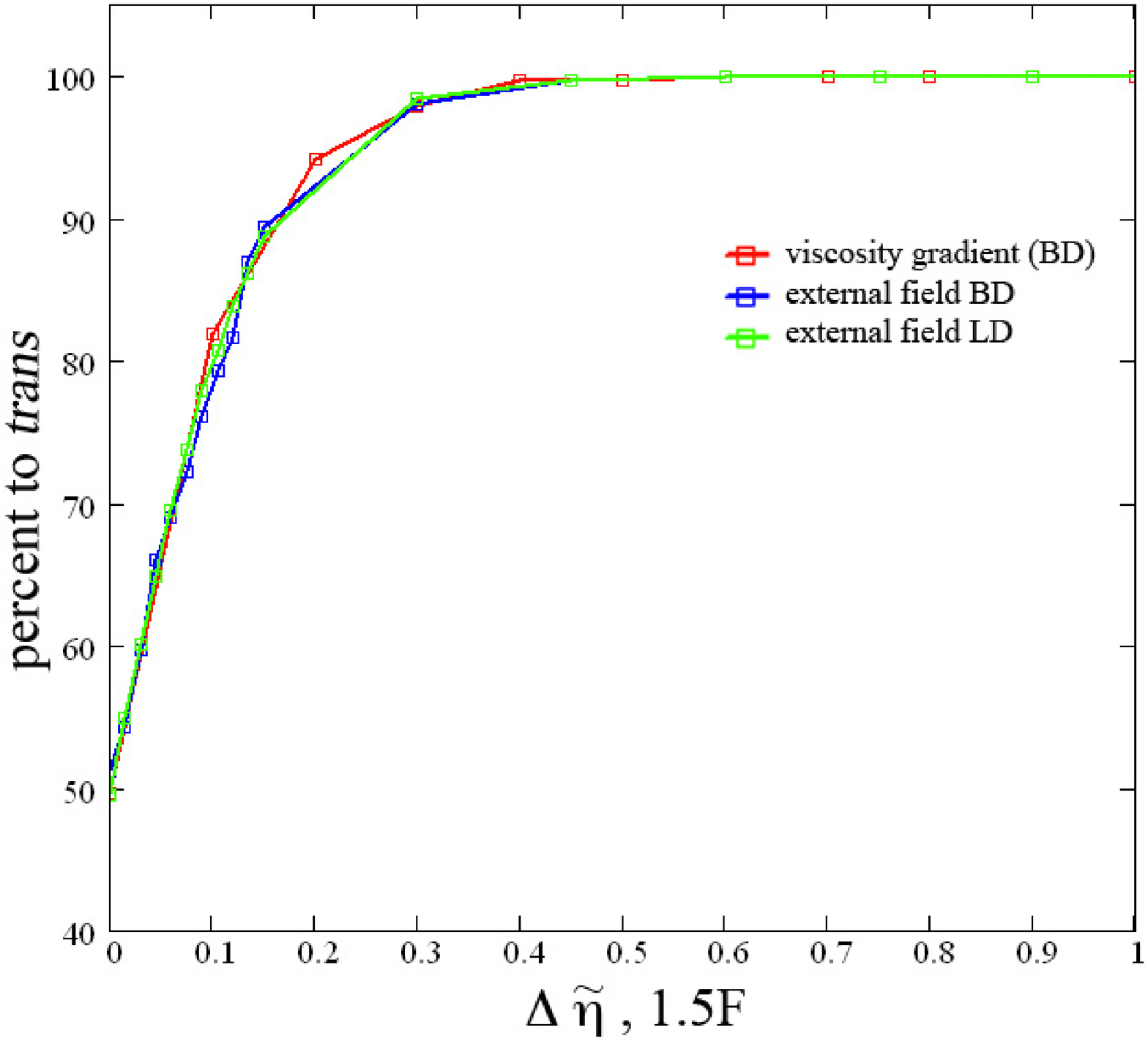} 
	\caption{Matching of the preferential direction for the BD viscosity gradient results (red, events to the high viscosity side),
	the external force model applied to BD simulations at a uniform viscosity (blue, events in the direction of the force),
	and the external force model applied to LD simulations at a uniform viscosity (green, events in the direction of the force).
	The polymer size is set to $N$=49.
	Good agreement is found taking $\Delta \tilde{\eta} = 1.5 F$.}
	\label{fig:EF_PD_BD}
\end{figure}

To measure the bias at the viscosity interface for both the LD and BD simulations, a simple force model is employed.
In this approach, the viscosity is held uniform at $\tilde{\eta}_C = \tilde{\eta}_T = 1$ and an external force is applied to the monomers at the interface by applying a force $F$ to any monomer which is in the pore.
We again focus on the $N$=49 case and start with the polymer halfway.
Performing these simulations for values of $F$ ranging from 0.01 to 10, the preferential direction and translocation time are recorded.
Figure \ref{fig:EF_PD_BD} displays the results for the preferential direction results for the force model applied to both LD and BD simulations along with the results for the BD viscosity gradient simulations.
In Fig. \ref{fig:EF_PD_BD}, there is excellent agreement between the external force simulations from LD and BD approaches.
This result demonstrates that, for a system of uniform viscosity, LD and BD simulations generally do give equivalent results
(although not shown, good agreement between the translocation times is also found with only a uniform rescaling of the times between the approaches being required).
Good agreement between the viscosity gradient case and the external force cases is obtained by rescaling the force such that $\Delta \tilde{\eta} = \tilde{\eta}_T - \tilde{\eta}_C = 1.5F$.
The preferential direction results thus allow us to estimate the magnitude of the bias at the interface induced by the viscosity gradient via $F = \Delta \tilde{\eta}/1.5$.
Considering that typical values for external fields in simulations examining driven translocation are in the range of 0.5 - 3 (in equivalent units),
this result indicates that the bias due to the viscosity gradient in BD simulations is of considerable magnitude.

\begin{figure}[h]
 	\centering
	\includegraphics[width=0.50\textwidth]{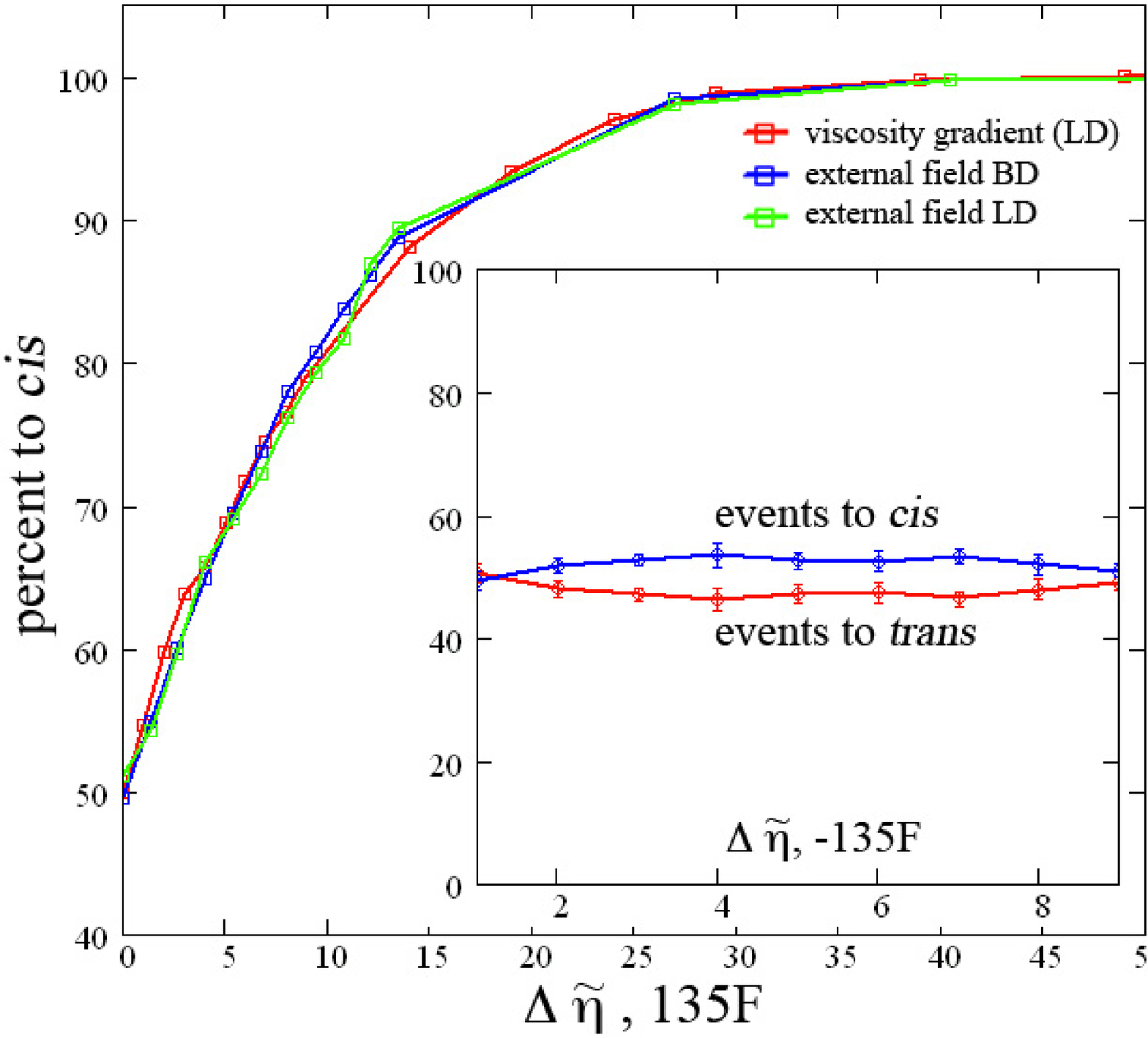} 
	\caption{Matching of the preferential direction for the LD viscosity gradient results (red, events to the low viscosity side),
	the external force model applied to BD simulations at a uniform viscosity (blue, events in the direction of the force),
	and the external force model applied to LD simulations at a uniform viscosity (green, events in the direction of the force).
	The polymer size is set to $N$=49.
	Good agreement is found taking $\Delta \tilde{\eta} = 135 F$.
	The inset shows the preferential direction when the external force and the bias due to viscosity gradient oppose each other.
	Using the $\Delta \tilde{\eta} = 135 F$ relation, the preferential direction is effectively nullified across the viscosity range studied.}
	\label{fig:EF_PD_LD}
\end{figure}
The same data plotted along with the LD viscosity gradient results is shown in Fig. \ref{fig:EF_PD_LD}.
Here, to get agreement between the external force and viscosity gradient results, the force must be rescaled by $\Delta \tilde{\eta} = 135F$ indicating that $F = \Delta \tilde{\eta}/135$.
The force induced by the viscosity gradient in LD simulations is much smaller than that typically employed to examine biased translocation.
Similarly, it is nearly two order of magnitude smaller from that generated in BD simulations.
This discrepancy also demonstrates how much more sensitive the BD simulations are to viscosity changes with a much stronger preferential direction being established at smaller values of $\Delta \tilde{\eta}$.
Note that the value of 135 seems to be independent of the polymer length $N$.
For simulations performed at shorter polymers ($N=25$) and longer polymers ($N=75$), using a factor of 135 was found to give good agreement between the viscosity gradient and force model results (data not shown).

The inset to Fig. \ref{fig:EF_PD_LD} displays the results from ``force balance" simulations in which the force is applied in the opposite direction to the effective force arising from the viscosity gradient.
Relating the external force to the viscosity gradient by $\Delta \tilde{\eta} = 135 F$, the forces effectively cancel each other out and the preferential direction all but vanishes.
This result further corroborates the external force model as well as the estimate of the magnitude of the bias due to the viscosity gradient.

Matching the preferential directions allows us to estimate the magnitude of the bias at the interface due to the viscosity difference.
In comparing the translocation times, there is another factor we must consider.
For the external force cases, an increasing $F$ necessarily decreases $\tau$.
However, in the case of the viscosity gradient, as discussed earlier, there are two factors to consider: 
i) the reduction in $\tau$ due to an increased bias and ii) an increase in $\tau$ as the average viscosity in the system is being increased.
To compare $\tau$ between the external force and viscosity gradient simulations, we must account for this latter factor.
To do so,  $\tau$ from the viscosity gradient simulations must be rescaled by a factor that depends on the viscosities across the membrane: $g(\tilde{\eta}_C,\tilde{\eta}_T)$.
We have recently shown that although the translocation time is independent of viscosity at very low $\tilde{\eta}$ values,
it decreases  as $1/\tilde{\eta}$ for intermediate to high viscosities (i.e., $\tilde{\eta} \ge 1$).
Using this result and recalling that in these simulations we fix $\tilde{\eta}_C$ and vary $\tilde{\eta}_T$, we assume the following form for the $g(\tilde{\eta}_C,\tilde{\eta}_T)$ scaling factor:
\begin{equation}
g(\tilde{\eta}_C, \tilde{\eta}_T) = A (\tilde{\eta}_T - \tilde{\eta}_C) + \tilde{\eta}_C.
\label{gfactor}
\end{equation}
This form captures the dependence on the viscosity difference $\Delta \tilde{\eta} = \tilde{\eta}_T - \tilde{\eta}_C$ and ensures that, 
in the limit where $\Delta \tilde{\eta} = 0$, the inversely proportional scaling of the translocation time with uniform viscosity given by $g = \tilde{\eta}_C = \tilde{\eta}_T$ is obtained.

\begin{figure}[h]
 	\centering
	\includegraphics[width=0.50\textwidth]{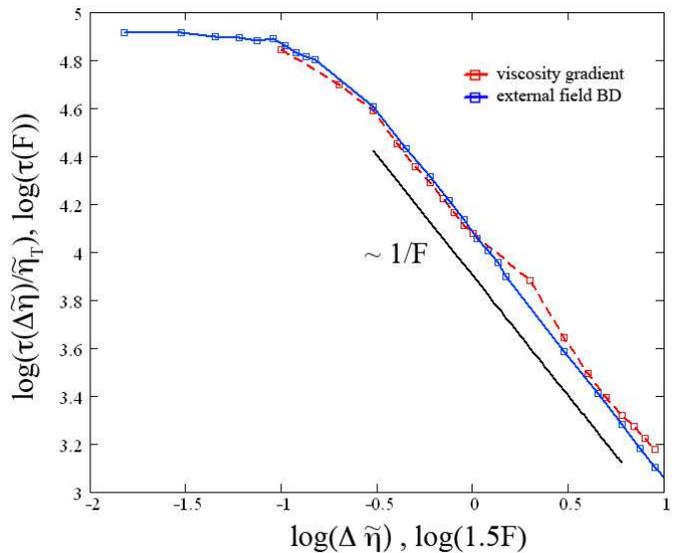} 
	\caption{Matching of the translocation times by plotting the BD viscosity gradient times normalized by $\eta_T$ along with the BD external field times.
	The polymer size is $N=49$.}
	\label{fig:EF_tau_BD}
\end{figure}

Beginning with the BD case, Fig. \ref{fig:EF_tau_BD} displays the translocation time corresponding to the external field model $\tau_E$ plotted against 1.5$F$.
Here we show the results from the driving force applied to the BD simulations such that $\tau_E(F=0) = \tau_{\Delta \tilde{\eta}}(\Delta \tilde{\eta} = 0 )$.
Fig. \ref{fig:EF_tau_BD} also shows the translocation time corresponding to the viscosity gradient results $\tau_{\Delta \tilde{\eta}}$ where the time has been normalized by $g$.
Good agreement is found when $A=1$ such that
\begin{eqnarray}
g & = & (\tilde{\eta}_T - \tilde{\eta}_C) + \tilde{\eta}_C \\
   & = & \tilde{\eta}_T.
\end{eqnarray}
This indicates that in BD where the events occur to the high viscosity side,
the rescaling of the translocation time to account for the changing viscosity in the system is achieved simply by normalizing $\tau$ by the viscosity on the high viscosity side.
This simple result demonstrates that as events occur towards this side, motion of the entire polymer through the high viscosity side is the rate limiting step for translocation.
Hence, for the BD results, good agreement for both the preferential direction and the translocation time 
is achieved by matching the external field to the viscosity gradient via $\Delta \tilde{\eta} = 1.5 F$ and normalizing the time scale of the dynamics by $\tilde{\eta}_T$.

Figure \ref{fig:EF_tau_BD} also demonstrates that there is a transition from an essentially $F$ independent region to a region where $\tau$ decreases as $1/F$ around $F^*=0.1/1.5=0.667$.
This crossover reflects the transition between the dynamics being determined by diffusive or driving processes 
and has been found before for studies examining driven translocation at low forces \cite{gauthier2008b}.
Below $F^*$, diffusion is dominant and, since the external force plays a minor role in determining the translocation time, $\tau$ is relatively independent of $F$.
Above this point, the driving force dominates and $\tau$ decreases proportional to $F$.
For the viscosity gradients studied, the BD results correspond to the high field limit again reflecting the large impact of the viscosity gradient in BD simulations.

\begin{figure}[h]
 	\centering
	\includegraphics[width=0.50\textwidth]{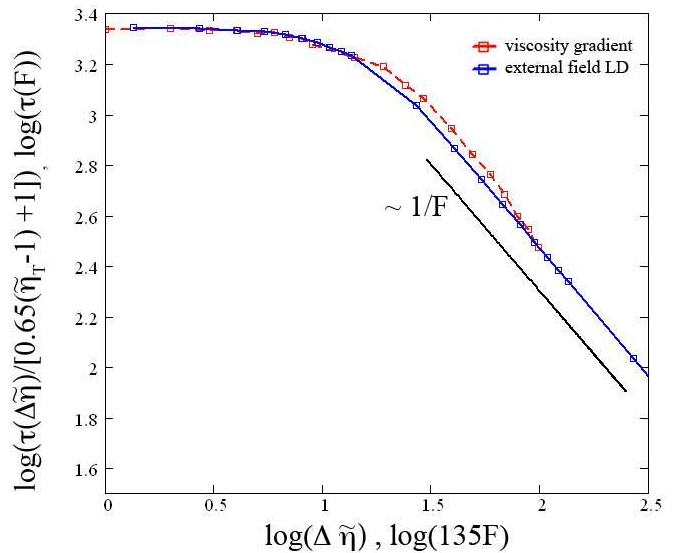} 
	\caption{Matching of the translocation times by plotting the LD viscosity gradient times normalized by $0.65(\tilde{\eta}_T-1)+1$ along with the LD external field times.
	The polymer size is $N=49$.}
	\label{fig:EF_tau_LD_log}
\end{figure}

For the LD case, $\tau_F$ plotted against 135$F$ and $\tau_\eta /g$ plotted against $\Delta \tilde{\eta}$ is shown in Fig. \ref{fig:EF_tau_LD_log}.
In this plot, the results for the external field generated from LD simulations are shown.
In the LD case, good agreement is found when $A=0.34$ such that
\begin{eqnarray}
g & = & 0.65(\tilde{\eta}_T - \tilde{\eta}_C) + \tilde{\eta}_C \\
   & = & 0.65(\tilde{\eta}_T - 1) + 1.
\end{eqnarray}
Contrary to the BD results where a direct scaling of $1/\tilde{\eta}_T$ was obtained, 
in LD there is a weaker dependence of the translocation time on the magnitude of the viscosity on the high viscosity side.
This result reflects that in LD, the events occur towards the low viscosity side and thus the impact of increasing the viscosity on the other side is reduced compared to BD.

Figure \ref{fig:EF_tau_LD_log} also shows that, contrary to the BD results, some of the LD results lie in the region where $\tau$ is relatively unaffected by $F$.
Again, the viscosity gradient results from LD correspond to low fields while from BD they correspond to high fields
(this is particularly true recalling that the range of $\Delta \tilde{\eta}$ simulated for BD was smaller than that for LD).

Hence, for both BD and LD, a force equivalent to the bias introduced by the viscosity gradient can be obtained.
For BD, the magnitude of the force is nearly 100 times larger than that for LD.
In fact, as the equivalent BD force is so large, the reduction in the translocation due to having a preferential direction 
is much greater than the increase in the translocation time due to having a higher average viscosity in the system.
Hence, the translocation time \textit{decreases} with an increasing $\tilde{\eta}_T$.
The converse is true in LD where the equivalent force is quite weak and the reduction in $\tau$ due to the preferential direction is negligible compared to the increase in the time scale of the dynamics due to increase $\tilde{\eta}_T$.
Correspondingly, $\tau$ increases as $\tilde{\eta}_T$ does.

\subsubsection{Physicality of the alogrithms}

The drastic differences between these results raise the question as to which one is physical.
We believe the LD result is the correct one for this scenario.
First, the equation of motion for LD is more correct in including the inertial term since  it does not require the assumption of overdamped dynamics implicit to BD.
Further, the results for the translocation time for the BD simulations seem to be unphysical.
It is difficult to believe that, 
given an average translocation time when $\tilde{\eta}_T = \tilde{\eta}_C = 1$, the translocation time is \textit{reduced} by a factor of four when the viscosity on the \textit{trans} side is doubled. 
As demonstrated by Fig. \ref{fig:BD_dyns}, the BD approach quickly yields a ratchet in which monomers jump from low to high viscosity and never cross back.
In fact, these dynamics result from the monomer jumping over the interface without feeling it 
such that a monomer from the low viscosity side is essentially artificially ``injected" into the high viscosity side without experiencing the change in viscosity.
As shown, this quickly suppresses diffusive motion at the interface, yields a rapid establishment of a 100\% preferential direction, and produces a dramatic drop in the translocation time.
These results are consistent with having a pump for monomers from the low viscosity side to the high viscosity side.
As there is no source for the energy required to drive such a pump, the BD results seem unphysical indicating that at least this implementation of BD is inappropriate for the system under study.

As the LD simulation results are taking to be correct for this scenario, the remainder of the paper focuses on the LD results.
However, it is worthwhile to bring attention to the remarkably different results obtained between LD and BD
as this example demonstrates that while these approaches generally yield the same result,
scenarios exist where the difference between these approaches can dictate the final result.

\subsection{One dimensional random walker in a viscosity gradient}

As mentioned, the LD results follow from a study we have published concerning a single particle at a viscosity interface \cite{dehaan2012c}.
In that work, the particle was initially placed at a sharp interface equidistant between two walls.
Here, we build on this approach to make it more analogous to the case of polymer translocation.
Comparison between the simplified model and the polymer results will then give insight into the role of the internal degrees of freedom of the polymer and the effect of the entropic barrier.
In the aforementioned study, Monte Carlo approaches for simulating the system were introduced and tested along with Langevin and Brownian Dynamics simulations.
While these approaches can be extended to the case of polymer translocation studied here,
in the current article we present results only from Langevin dynamics simulations for the sake of simplicity of comparison to the polymer results.

To study translocation, several studies (including the earliest theoretical treatments of the problem) have used a quasistatic approximation to reduce the process to a one dimensional random walker (1DRW) 
\cite{sung1996,muthukumar1999,chuang2001,gauthier2008,hassanabad2009,dehaan2011}.
The basic idea is that, if the polymer can be considered in equilibrium at all times, translocation can be represented as a single random walker traversing an entropic barrier which arises from confining the polymer within the nanopore.
In a similar fashion, the effects of the viscosity gradient can be included in this model.
Considering a polymer halfway through the pore, there will be $N/2$ monomers on \textit{cis} experiencing $\tilde{\eta}_C$ and $N/2$ monomers on \textit{trans} experiencing $\tilde{\eta}_T$.
The average friction coefficient for this configuration is thus $(\tilde{\eta}_C+\tilde{\eta}_T)/2$.
Defining a translocation coordinate $s$ which indicates how many monomers are on the \textit{trans} side, this definition can be extended:
\begin{equation}
\tilde{\eta}(s) = \frac{(N-s)\tilde{\eta}_C + s\tilde{\eta}_T}{N}
\end{equation}
where $N$, being the polymer length, corresponds to the separation between the absorbing walls for the 1DRW.

Using this form, Langevin dynamics simulations for a one dimensional random walker in a linear viscosity gradient were performed.
Simulations were done for both a ``free" particle (subject only to the viscosity gradient) and a particle traversing an entropic barrier.
The form for the entropic barrier is given by the quasistatic approximation \cite{sung1996,muthukumar1999,chuang2001,gauthier2008,hassanabad2009,dehaan2011}:
\begin{eqnarray}
U\left(\frac{s}{N}\right) & = & -kT\ln \left( \frac{1}{\left(1-\left(\frac{s}{N}\right)\right)^{1-\gamma}} \frac{1}{\left(\frac{s}{N}\right)^{1-\gamma}}\right)
\end{eqnarray}
where $\gamma$ is the surface exponent set to 0.69 here \cite{eisenriegler1982}
such that a force given by
\begin{eqnarray}
\frac{F\left(\frac{s}{N}\right)}{kT} & = & \frac{1}{N}\left(\frac{1-\gamma}{1-\frac{s}{N}} - \frac{1-\gamma}{\frac{s}{N}}\right)
\label{Fkt}
\end{eqnarray}
is applied to the particle.

\begin{figure}[h]
 	\centering
	\includegraphics[width=0.50\textwidth]{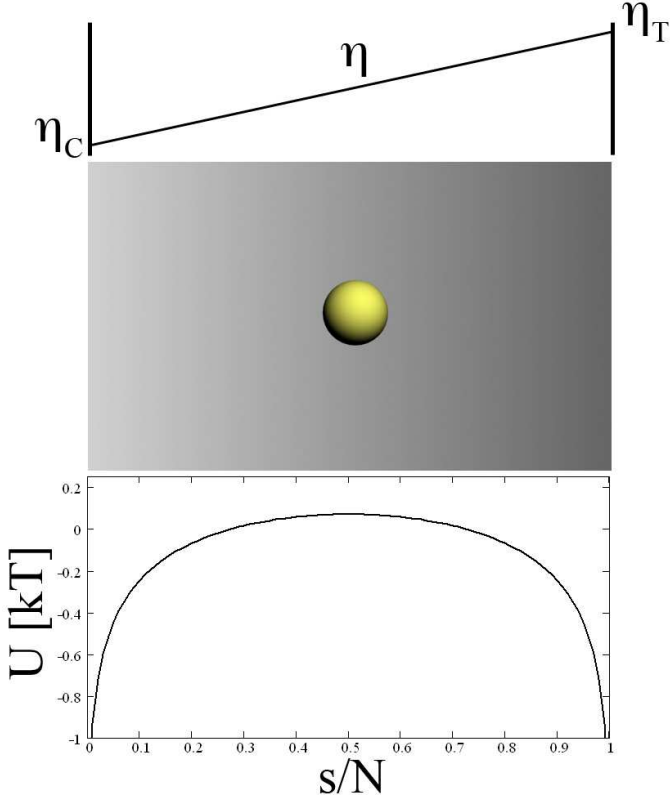} 
	\caption{Schematic of 1D random walker in a viscosity gradient}
	\label{fig:1D_scheme}
\end{figure}

\begin{figure}[h]
 	\centering
	\includegraphics[width=0.50\textwidth]{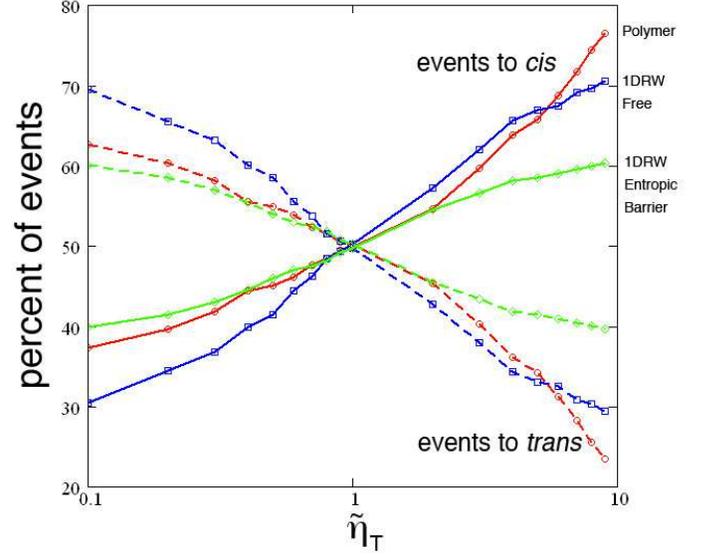} 
	\caption{Percent of events absorbed at the \textit{cis} wall (solid) and the \textit{trans} wall (dashed) for the polymer simulations (red),
	the 1DRW viscosity gradient simulations (blue),
	and the 1DRW simulations including the entropic barrier (green).
	The polymer size is set to $N$=49.}
	\label{fig:1D_PD}
\end{figure}

As shown in Fig. \ref{fig:1D_scheme}, the particle was initially placed in between two absorbing walls.
Simulations were performed for $\tilde{\eta}_T = 0.1$ to 9.0 with $\tilde{\eta}_C$ held at 1 where $\tilde{\eta}_T$ is now the viscosity at the right wall and $\tilde{\eta}_C$ is the viscosity at the left wall.
The percent of events absorbed at either wall and the mean first passage time (MFPT) were recorded.
The results for the preferential direction are shown in Fig. \ref{fig:1D_PD} along with the polymer results.
In agreement with the LD polymer results, the particle preferentially ends up at the low viscosity wall.

Comparison between the 1DRW and polymer results indicate several things.
First, the consistent establishment of a preferential direction to the low side for all three simulation approaches 
demonstrates that the internal degrees of freedom of the polymer are not needed to obtain this result;
a preferential direction can be established for just a single particle with or without the entropic barrier (this is also true for the case of a sharp interface \cite{dehaan2012c}).

However, the internal degrees of freedom of the polymer do have an affect on the dynamics.
Examination of the preferential direction plotted logarithmically in $\tilde{\eta}$ shows that the results for the 1DRW are symmetric: $n_T/n_C(\tilde{\eta}_T) = n_C/n_T(1/\tilde{\eta}_T)$.
Conversely, the results for the polymer translocation are clearly not symmetric.
The difference between these two cases is that for the polymer, the internal degrees of freedom introduce another time scale into the system: the time required for the polymer to relax.
The relaxation time breaks the symmetry between $\tilde{\eta}_C/\tilde{\eta}_T$ and $\tilde{\eta}_T/\tilde{\eta}_C$.
We have recently published a study examining the unbiased translocation dynamics as a function of viscosity.
These results indicate that, for $N$=49, the dynamics at $\tilde{\eta}=0.1$ are essentially quasistatic while for $\tilde{\eta}=1$ and higher, nonequilibrium effects would be expected.
Hence, at $\tilde{\eta}_T/\tilde{\eta}_C = 0.1$, the dynamics on $\textit{trans}$ are quasistatic while those on $\textit{cis}$ are not.
However, at $\tilde{\eta}_T/\tilde{\eta}_C = 10$, neither the dynamics on $\textit{trans}$ or $\textit{cis}$ are expected to be quasistatic.
The result is the asymmetric results shown in Fig. \ref{fig:1D_PD}.

Interestingly, the nonequilibrium effects appear to be strengthening the preferential direction since $n_C(\tilde{\eta}_T) > n_T (1/\tilde{\eta}_T)$.
To give context to this, consider a scenario in which the viscosity on both sides was low enough to yield quasisatic dynamics.
For this case, the preferential direction would arise from the bias felt by monomers at the interface towards the low viscosity side.
If the viscosity on both sides is now raised (while maintaing the same ratio) 
such that neither side corresponds to quasistatic dynamics, another bias is introduced due to the differing relaxation times across the interface.
Given that the relaxation time will be lower on the low viscosity side, 
movement of monomers towards the low viscosity side will in general be more successful than attempts to move monomers to the high viscosity side
(for a detailed study of the effects of viscosity on memory effects during translocation, see \cite{dehaan2012a}).
This extra bias resulting from this discrepancy of relaxation times adds to the tendency for monomers towards the low viscosity side and thus yields a stronger preferential direction.

Comparing the 1DRW results, a noticeable drop in the strength of the preferential direction is observed when the effects of the entropic barrier are included.
For this case of a continuous viscosity gradient, one can consider the particle to be experiencing a bias to the low viscosity wall at all locations.
In the absence of an entropic barrier, the net effect of this bias yields a strong preferential direction.
When the entropic barrier is added, this bias still dominates near the middle of the system where the entropic barrier is flat and thus the resulting force is negligible.
However, when the particle is quite near to a wall, the entropic barrier yields a large force towards that wall.
In these regions, the force arising from the viscosity gradient will be negligible.
The particle thus effectively sees a reduced viscosity gradient corresponding to the viscosity drop over the middle of the system.
As shown previously, 
a reduced viscosity gradient yields a weaker preferential direction and thus there is a drop in the strength of the preferential direction on including the viscosity gradient in the simulations.
This result will become important when considering the results for different polymer lengths discussed below.

\begin{figure}[h]
 	\centering
	\includegraphics[width=0.50\textwidth]{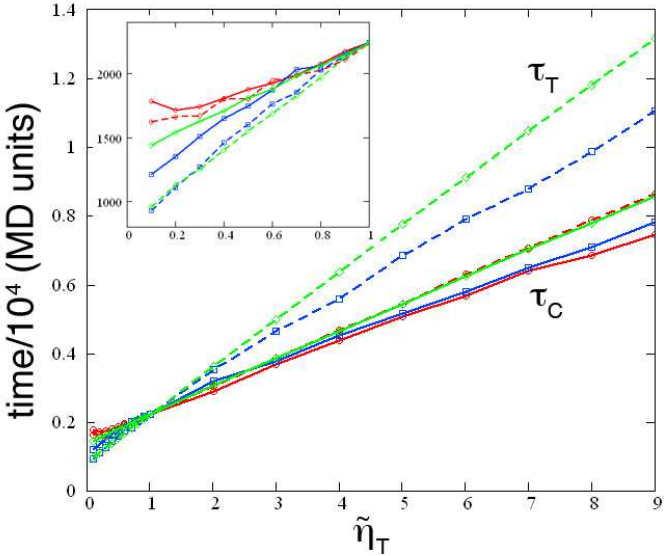} 
	\caption{Mean first passage time to \textit{cis} (solid lines) and \textit{trans} (dasehd lines) for polymer translocation (red), a free 1DRW (blue), and a 1DRW traversing an entropic barrier (green).
	For ease of comparison, all times are normalized to be equal at $\tilde{\eta}_C=\tilde{\eta}_T$.}
	\label{fig:1D_times}
\end{figure}

An additional bias to the low viscosity side for higher $\tilde{\eta}_T$ may also explain the discrepancies for the times shown in Fig. \ref{fig:1D_times}.
When the times are normalized to be equal at $\tilde{\eta}_C=\tilde{\eta}_T$, the 1DRW MFPTs overestimate the polymer translocation times for $\tilde{\eta}_T > 1$.
This result is consistent with the polymer experiences an additional bias towards the low viscosity side such that $\tau$ does not grow as fast with $\tilde{\eta}_T$ as for the 1DRW simulations.
1
In contrast, the 1DRW results underestimate the translocation time when $\tilde{\eta}_T < 1$ (inset to Fig. \ref{fig:1D_times}).
In reducing the problem to a one dimensional process, many aspects of translocation are neglected.
In the 1DRW model, lower viscosity values directly correspond to higher diffusion coefficients and thus $\tau$ continues to decrease with decreasing $\eta_T$.
On the other hand, for a polymer in three dimensions, no matter how low the viscosity, the polymer must thread through the constricting nanopore in order to translocate.
In fact, we have recently shown that at very low (uniform) viscosities, the translocation time is independent of the viscosity;
friction with the pore is entirely dominant and lowering the viscosity does not speed up the process \cite{dehaan2012a}.
Friction with the pore can thus limit the reduction of the translocation time and as this effect is missing in the 1DRW model, lower MFPTs are predicted.

\subsection{Scaling with $N$}

\begin{figure}[h]
 	\centering
	\includegraphics[width=0.50\textwidth]{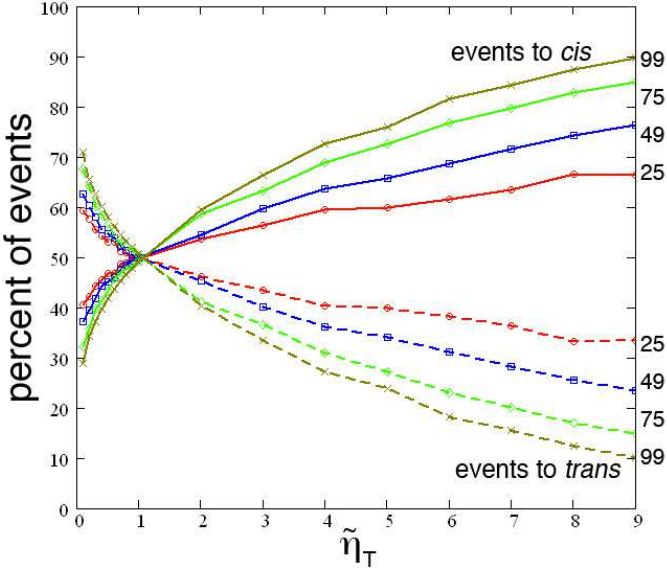} 
	\caption{Percent of events towards \textit{cis} (solid) and \textit{trans} (dashed) wall for polymers of size $N=25$ (red), $N=49$ (blue), $N=75$ (green), and $N=99$ (brown) as function of $\tilde{\eta}_T$.}
	\label{fig:scale_PD}
\end{figure}

\begin{figure}[h]
 	\centering
	\includegraphics[width=0.50\textwidth]{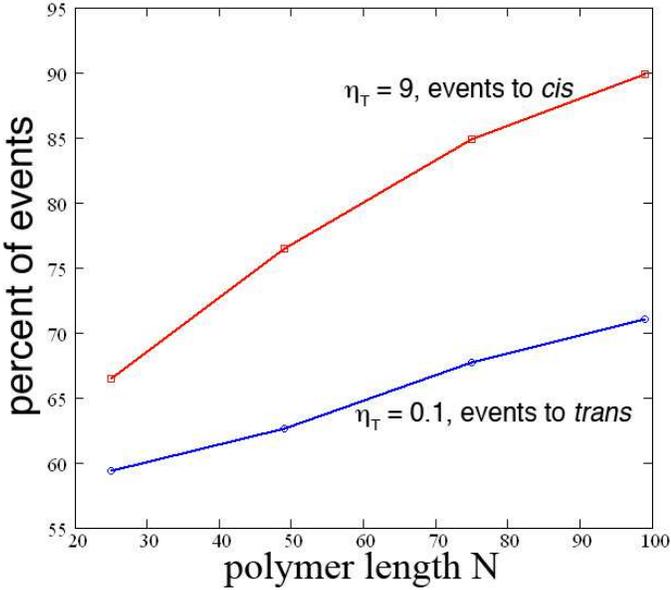} 
	\caption{Saturation of the preferential direction as a function of the polymer size $N$ for the two extreme viscosity gradient cases:
	$\tilde{\eta}_T=9$, $\tilde{\eta}_C=1$ (events towards $cis$) shown in red and $\tilde{\eta}_T=0.1$, $\tilde{\eta}_C=1$ (events towards $trans$) shown in blue.}
	\label{fig:scale_satur}
\end{figure}

While the above discussion concerns only the results for $N$=49, simulations were also performed for $N = 25, 75, 99$.
The results for the preferential direction are shown in Fig. \ref{fig:scale_PD}.
The strength of the preferential direction consistently increases with increasing molecular weight.
This increase is highlighted in Fig. \ref{fig:scale_satur} where the preferential direction is plotted against the polymer length for the two extreme viscosity gradient cases:
$\tilde{\eta}_T=9$, $\tilde{\eta}_C=1$ and $\tilde{\eta}_T=0.1$, $\tilde{\eta}_C=1$.
For the $\tilde{\eta}_T=9$, the number of events towards $cis$ increases to near 90\% at $N=99$.
Correspondingly, the saturation of the preferential direction is evident as the curve must eventually plateau at 100 \%.
For the $\tilde{\eta}_T=0.1$ data, the increase in the strength of the preferential direction is more modest and the saturation is not evident over the polymer length range studied.
Comparing between the curves, the disparity between the curves for viscosity ratios which are nearly the inverse of each other again demonstrates the asymmetry of the dynamics intrinsic to the polymer results.

Considering the increase in the preferential direction with $N$, 
we note that the preferential direction for the 1DRW is independent of the separation of the walls (results are not shown).
Beyond trivially increasing the MFPTs, there is no effect of increasing the separation between the walls.
Recalling that the internal degrees of freedom of the polymer yield an additional bias towards the low viscosity side,
this strengthening of the preferential direction with increasing $N$ can be explained by noting that this extra bias will increase with $N$.
That is, as the polymer grows, the disparity in relaxation times between the \text{cis} and text{trans} side grows yielding a stronger bias to the low viscosity side and thus a stronger preferential direction.

There is another effect to consider here.
Recall that adding the entropic barrier to the 1DRW model weakened the preferential direction.
As the particle/polymer nears translocation, entropic effects dominate the bias due to the viscosity gradient and a reduced effect of the gradient is observed.
It has been shown that for polymer translocation at viscosities where the process is not quasistatic, the effects of the entropic barrier diminish with length \cite{dehaan2012b}.
That is, entropic effects play a bigger role at $N$=25 than $N$=99.
Correspondingly, the diminishment of the viscosity gradient bias due to the entropic barrier is reduced as $N$ increases and the strength of the preferential direction increases with $N$.

Hence, both the growing discrepancy between relaxation times on either side of the pore 
and the postulate that the polymer experiences a large portion of the viscosity gradient result in a preferential direction that grows with $N$.
While it may be possible to design tests to unravel these effects,
for now we simply note that there are two possible mechanisms - both arising from the internal degrees of freedom of the polymer - which could explain this result.

\begin{figure}[h]
 	\centering
	\includegraphics[width=0.50\textwidth]{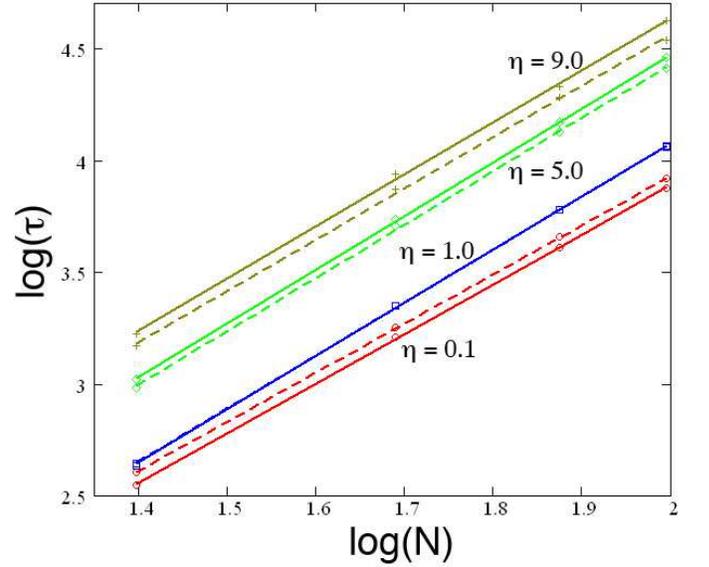} 
	\caption{Translocation times to \textit{trans} (solid lines) and \textit{cis} (dashed lines) for $\tilde{\eta}_C$ = 1.0 and $\tilde{\eta}_T$ = 0.1 (red), 1.0 (blue), 5.0 (green), and 9.0 (brown) as a function of $N$. }
	\label{fig:scale_times}
\end{figure}

The scaling of the translocation times was also examined.
Figure \ref{fig:scale_times} displays the translocation to both the \textit{trans} and \textit{cis} sides for $N=25, 49, 75, 99$ at 4 different values of the viscosity on the \textit{trans} side.
Note that for $\tilde{\eta}_T < \tilde{\eta}_C$, $\tau_T < \tau_C$ and that for $\tilde{\eta}_T > \tilde{\eta}_C$, $\tau_T > \tau_C$.
For $\tilde{\eta}_T = \tilde{\eta}_C$, $\tau_T \approx \tau_C$ as required.

\begin{figure}[h]
	\centering
	\includegraphics[width=0.50\textwidth]{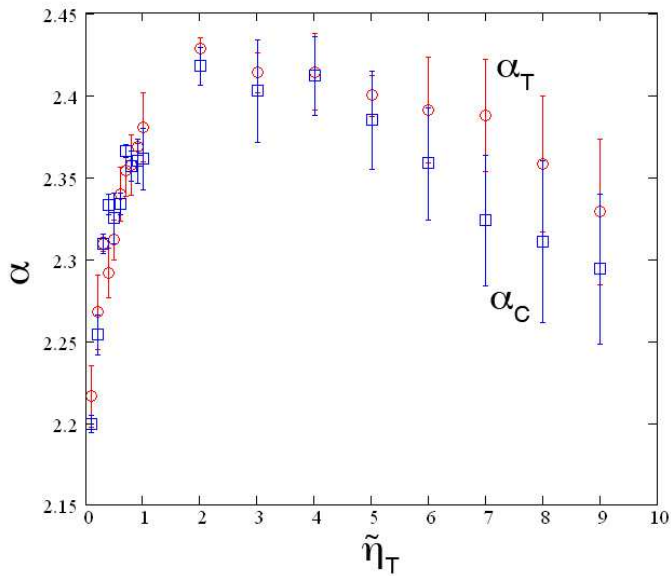} 
	\caption{Scaling exponent $\alpha$ for trajectories occurring towards \textit{trans} (red) and \textit{cis} (blue) as a function of $\tilde{\eta}_T$.}
	\label{fig:scale_alpha}
\end{figure}

Assuming a scaling of translocation time with polymer length given by
\begin{equation}
\tau \sim N^{\alpha},
\end{equation}
values for the $\alpha$ exponent can be obtained from the slopes of the lines in Fig. \ref{fig:scale_times}.
These results are shown in Fig. \ref{fig:scale_alpha}.
There is a maximum just beyond $\eta_T =1.0$.
On either side of this region, the scaling decreases with an increasing difference in viscosity.

The physical picture here is that, if translocation is predominantly a diffusive process, one would expect $\tau \sim N^{2}$
(in fact, due to nonequilibrium effects, values of $\alpha$ greater than 2 are consistently obtained).
On the other hand, in the limit where the bias is high enough to completely suppress diffusion, one would expect $\tau \sim N$.
Thus, the introduction of a bias reduces the $\alpha$ exponent.
Considering that the viscosity gradient introduces a bias that grows with an increasing difference, 
$\alpha$ decreases both as $\tau_T/\tau_C \rightarrow 0$ and as $\tau_T/\tau_C \rightarrow \infty$.

However, the maximum is not at $\tau_T/\tau_C = 1$ as this argument would imply.
Rather, it is just to the right of this point.
As mentioned, we recently studied the effect of viscosity on the unbiased translocation dynamics \cite{dehaan2012a, dehaan2012b}.
It was found that the $\alpha$ exponent increased significantly with increasing viscosity due to varying impacts of nonequilibrium effects.
This effect of an increased $\alpha$ at higher $\tilde{\eta}$ thus counters the decrease in $\alpha$ as $\tau_T/\tau_C \rightarrow \infty$.

From the data, there is a crossover between which effect is dominant:
for points just beyond $\tau_T/\tau_C = 1$, $\alpha$ increases due to the higher viscosity on $\textit{trans}$;
beyond $\tau_T/\tau_C = 2$, $\alpha$ decreases due to the higher \textit{difference} between $\tilde{\eta}_T$ and $\tilde{\eta}_C$.

\section{Conclusion}

In this work, we examine the translocation of a polymer through a nanopore when the viscosity differs across the membrane.
To do so, the viscosity on $cis$ is fixed at 1.0 and simulations are performed for $trans$ values varying from 0.1 to 9.
Interestingly, the results were found to change drastically between simulation algorithms.
For the establishment of a preferential direction, the polymer tends to the low viscosity side in LD while in BD, more events occur towards the high viscosity side.
For the translocation time, $\tau$ varies essentially linearly with the $trans$ viscosity in LD.
In BD, $\tau$ decreases with an increasing viscosity difference across the pore such that $\tau$ is a maximum when the viscosity is uniform.

These conflicting results provide an instructive example of the care which must be taken in choosing a simulation algorithm when studying a particular system.
For the study of polymer dynamics, LD and BD generally produce equivalent results and, for studies of polymer translocation, both have been used extensively.
In all such cases, the viscosity in the system was uniform and thus the results were insensitive to the choice of algorithm.
In fact, in this work for the simple force model simulations preformed with a uniform viscosity, BD and LD were shown to produce equivalent results.
However, if we now consider a system where the viscosity on $trans$ is twice that on $cis$, the answer of where a polymer that starts halfway will end up turns out to depend on the simulation approach:
in BD the polymer ends up on the ``thick" side but in LD the polymer ends up on the ``thin" side.

To delve into these results, the details of the dynamics and magnitude of the effective bias were studied.
In BD, increasing the viscosity on $trans$ even a relatively small amount corresponds to a considerable force at the interface.
Correspondingly, the driving of monomers across the interface quickly suppresses all diffusive motion and the translocation time decreases rapidly.
In LD, the bias introduced from a viscosity gradient is nearly two orders of magnitude smaller than that produced in BD.
Likewise, the diminishment of diffusive motion is much weaker and the translocation time increases with an increasing viscosity on $trans$.
From this comparison, we believe that, in general, the LD results are correct.
Fundamentally, the difference arises from the inertial term in the equation of motion in LD which allows monomers to sense the interface as they pass over it.
In BD where there is no inertial term, the dynamics occur entirely according to the viscosity at the initial point of the jump.
Thus, not only is the former more physically complete, but the results are more reconcilable with physical intuition;
it is hard to imagine that doubling the viscosity on $trans$ will result in all events occurring to this side in 1/4 the time.

Finally, the scaling aspects of these results were examined.
We find that the strength of the preferential direction increases with increasing polymer length.
Comparison to a single random walker at a viscosity interface indicates that this $N$ dependence arise from the internal degrees of freedom of the polymer.
For the scaling of the translocation time with $N$, the $\alpha$ exponent generally decreases as the viscosity gradient grows.
This trend is expected as the process moves from diffusive to driven.
However, the maximum $\alpha$ does not occur when the viscosity is uniform.
Rather, if the viscosity on $trans$ is increased from one to two, $\alpha$ increases.
This result is consistent with previous observations that, for a system of uniform viscosity, $\alpha$ increases with increasing viscosity.
There is an interplay between these two effects and thus $\alpha$ increases for small increases in the viscosity on $trans$ but decreases as the viscosity gradient grows to larger values.

The translocation of a polymer across a viscosity gradient thus presents a wealth of results.
Not only does this system represent a clear case where BD and LD yield very different results,
but the results themselves present implications for naturally occurring cases of translocation.
Taking the LD results to be more physical,
a polymer that is halfway between a thin and thick medium will experience a bias driving it towards the thin side.
The strength of this bias increases not only with the viscosity gradient, but also with the length of the polymer.
Likewise, the scaling of the translocation time with the polymer length is affected with lower values of $\alpha$ being found for larger gradients.
Although the simulations indicate that this bias is a relatively small force,
these effects are likely to arise in both natural and synthetic cases of translocation
and thus may be of interest both for reconciling experimental data with theoretical predictions
and for guiding the design of nanofluidic devices.

\section{Acknowledgements}

Simulations were performed using the ESPResSo package \cite{limbach2006} on the SHARCNET computer system (www.sharcnet.ca) using VMD \cite{hump1996} for visualization.
This work was funded by the NSERC and the University of Ottawa.


\begin{thebibliography}{99}

\bibitem{albe89} B. Alberts, D. Bray, J. Lewis, M. Raff, K. Roberts, and J.D. Watson, \emph{Molecular Biology of the Cell} (Garland Publishing, New York, 1989).
\bibitem{black89} L.W. Black, \emph{Annu. Rev. Microbiol.}, \textbf{43}, 267-92, (1989).
\bibitem{rapoport2007} T.A. Rapoport, \emph{Nature}, \textbf{450(29)}, 663-669, (2007).
\bibitem{branton2008} D. Branton, D. W. Deamer, A. Marziali, H. Bayley, S. A. Benner, T. Butler, M. Di Ventra, S. Garaj, A. Hibbs, X. Huang, S. B. Jovanovich, P. S. Krstic, S. Lindsay, X. S. Ling, C. H. Mastrangelo, A. Meller, J. S. Oliver, Y. V. Pershin, J. M. Ramsey, R. Riehn, G. V. Soni, V. Tabard-Cossa, M. Wanunu, M. Wiggin, and J. A. Schloss, \emph{Nat. Biotechnol.}, \textbf{26}, 1146Ð53, (2008).
\bibitem{muthubook} M. Muthukumar, \emph{Polymer Translocation} (CRC Press, New York, 2011).
\bibitem{wie2007} D. Wie, W. Yang, X. Jin and Q. Liao, \emph{J. Chem. Phys.}, \textbf{126}, 204901, (2007).
\bibitem{kapahnke2010} F. Kapahnke, U. Schmidt, D.W. Heermann, and M. Weiss, \emph{J. Chem. Phys.}, \textbf{132}, 164904, (2010). 
\bibitem{gopinathan2007} A. Gopinathan and Y.W. Kim. \emph{Phys. Rev. Lett.}, \textbf{99}, 228106, (2007).
\bibitem{ambjornsson2004} T. Ambj\"{o}rnsson and R. Metzler, \emph{Phys. Biol.}, \textbf{1(2)}, 77-88 (2004).
\bibitem{yu2011} W. Yu and K. Luo, \emph{JACS}, \textbf{133}, 13565-13570 (2011).
\bibitem{park1998} P.J. Park and W. Sung, \emph{J. Chem. Phys.}, \textbf{108(7)}, 3013-3018, (1998).
\bibitem{milchev2004} A. Milchev and K. Binder, \emph{J. Chem. Phys.}, \textbf{121(12)}, 6042-6051, (2004).

\bibitem{slater2009} G.W. Slater, C. Holm, M.V. Chubynsky, H.W. de Haan, A. Dub\'{e}, K. Grass, O. Hickey, C. Kingsburry, D. Sean, T.N. Shendruk, and L. Zhan, \emph{Electrophoresis}, \textbf{30(5)}, 792-818, (2009). 
\bibitem{weeks1978} J.D. Weeks, D. Chandler, H.C. Anderson. \emph{J. Chem. Phys.}, \textbf{54}, 5237-5247, (1978).
\bibitem{grest1986} G.S. Grest and K. Kremer. \emph{Phys. Rev. A.}, \textbf{33}, 3628-3631, (1986).
\bibitem{dehaan2010} H.W. de Haan and G.W. Slater, \emph{Phys. Rev. E}, \textbf{81}, 051802 (2010).

\bibitem{dehaan2012c} H.W. de Haan, M.V. Chubynsky, and G.W. Slater, http://arxiv.org/abs/1208.5081, \emph{submitted for publication, Phys. Rev. E}.
\bibitem{lancon2002} P. Lan\c{c}on, G. Batrouni, L. Lobry, N. Ostrowsky, \emph{Physica A},\textbf{304} 65-76 (2002).
\bibitem{sokolov2010} I.M. Sokolov, \emph{Chem. Phys.}, \textbf(375), 359-363, (2010).

\bibitem{sung1996} W. Sung and P.J. Park, \emph{Phys. Rev. Lett.}, \textbf{77(4)}, 783-786, (1996).
\bibitem{muthukumar1999} M. Muthukumar, \emph{J. Chem. Phys.}, \textbf{111(22)}, 10371-10374, (1999).
\bibitem{chuang2001} J. Chuang, Y. Kantor, M. Kardar, \emph{Phys. Rev. E}, \textbf{65}, 011802, (2001).
\bibitem{gauthier2008} M.G. Gauthier and G.W. Slater, \emph{J. Chem. Phys.}, \textbf{128}, 0065103, (2008).
\bibitem{hassanabad2009} M.F. Hassanabad and J.M. Polson, \emph{Physics in Canada}, \textbf{65(3)}, 126-129, 2009.
\bibitem{dehaan2011} H.W. de Haan and G.W. Slater, \emph{J. Chem. Phys.}, \textbf{134}, 154905 (2011).
\bibitem{eisenriegler1982} E. Eisenriegler, K. Kremer, and K. Binder, \emph{J. Chem. Phys.}, \textbf{77}, 6296-6320 (1982).

\bibitem{dehaan2012a} H.W. de Haan and G.W. Slater, \emph{J. Chem. Phys.}, \textbf{136}, 154903 (2012).
\bibitem{dehaan2012b} H.W. de Haan and G.W. Slater, \emph{J. Chem. Phys.}, \textbf{136}, 204902 (2012).

\bibitem{gauthier2008b} M.G. Gauthier and G.W. Slater, \emph{J. Chem. Phys.}, \textbf{128}, 205103 (2008).

\bibitem{limbach2006} H.J. Limbach, A. Arnold, B.A. Mann, C. Holm. \emph{Comput. Phys. Commun.}, \textbf{174}, 704-727, (2006).

\bibitem{hump1996} W. Humphrey, A. Dalke, and K. Schulten, \emph{J. Molec. Graphics} \textbf{14}, 33-38 (1996).

\end{thebibliography}
\end{document}